\documentclass[twocolumn,aps,showpacs,superscriptaddress,longbibliography]{revtex4-1}
\usepackage{graphicx}
\usepackage{latexsym}

\begin{document}

\title{Reconstruction of joint photon-number distributions of twin beams incorporating spatial noise reduction}

\author{Jan {Pe\v{r}ina,~Jr.}}
\email{jan.perina.jr@upol.cz}
\affiliation{Joint Laboratory of Optics of Palack\'{y} University and
Institute of Physics of CAS, Faculty of Science, Palack\'{y}
University, 17. listopadu 12, 77146 Olomouc, Czech Republic}
\author{V\'{a}clav Mich\'{a}lek}
\affiliation{Joint Laboratory of Optics of Palack\'{y} University and
Institute of Physics of CAS, Faculty of Science, Palack\'{y}
University, 17. listopadu 12, 77146 Olomouc, Czech Republic}
\author{Ond\v{r}ej Haderka}
\affiliation{Joint Laboratory of Optics of Palack\'{y} University and
Institute of Physics of CAS, Faculty of Science, Palack\'{y}
University, 17. listopadu 12, 77146 Olomouc, Czech Republic}

\begin{abstract}
A method for reconstructing joint photon-number distributions of twin beams from the experimental
photocount histograms is suggested and experimentally implemented. Contrary to the standard reconstruction methods,
it incorporates spatial noise reduction based on spatial pairing of photons. Superior performance
of the method above the usual one for the maximum-likelihood approach is demonstrated.
\end{abstract}


\maketitle

\section{Introduction}

In classical optics, an optical field is characterized by its
intensity and phase that is determined in an optical
interferometer and is given with respect to certain reference phase
\cite{Born1999}. However, as the formulation of quantum theory of
coherence \cite{Glauber1963} revealed, the measurement of
intensity represents only a certain limiting case appropriate for
the characterization of intense optical fields. For weak optical
fields composed of individual photons, the determination of a
complete photocount (i.e. photo-electron) distribution is
necessary \cite{Mandel1995}. The Mandel detection formula
\cite{Mandel1995} derived in the early days of quantum optics then
provides the bridge between the measured photocount
distribution and the distribution of (integrated) intensity
that describes the analyzed field \cite{Perina1991}.

Investigations of the detection of weak optical fields brought
considerable attention to the role of noises present not only in
the detection process but also in the observed optical fields
\cite{Saleh1978,Perina1991}. The noise of a detector, that is
composed of the intrinsic quantum shot-noise and an additional
electronic noise, can be independently quantified and subsequently
removed from the experimental data, at least in principle. Contrary to this, the
optical noise superimposed to the observed optical field during
its propagation to the detector cannot be usually eliminated and
such noise is considered as a part of the optical field. However,
this is not the case of twin beams that are composed of photon
pairs \cite{Jedrkiewicz2004,Bondani2007,Blanchet2008,Brida2009a}.
The optical noise in a twin beam can be indirectly identified by
exploiting spatial correlations of photons in a photon pair that
naturally emerge during the generation of a twin beam
\cite{Jost1998,Joobeur1994,Joobeur1996,Brambilla2004,Brambilla2010,Blanchet2008,Fedorov2008,Hamar2010,Tasca2013,Fickler2013,Fickler2014,Jachura2015}.
Such reduction of the noise may be useful whenever twin beams are
applied, for instance in quantum metrology
\cite{Lloyd2008,Giovannetti2011}, quantum imaging
\cite{Pittman1995,Gatti2008,Brida2009b,Tasca2013} or quantum
communications \cite{Saleh1987}.

Spatial correlations of photons in a twin beam are described by
the intensity cross-correlation function that allows to define a
correlated area \cite{Hamar2010}. When an idler photon is detected
inside the correlated area belonging to an already registered
signal photon, both photons are considered as members of a single
photon pair. This fact can be used to reduce the amount of noise
present in the joint signal-idler photocount distribution obtained
after the detection of a twin beam \cite{PerinaJr2017b}. The
strength of such noise reduction depends on the extension of a
detection area in the idler beam, in which we expect the idler
photon accompanying an already registered signal photon. The
smaller the detection area is, the more efficient the noise
reduction is. However, when the detection area is becoming smaller
than the correlated area some idler photons do not fall into the detection area
and so both photons from a photon pair cannot be simultaneously detected
('photon pairs are statistically partially broken'). Under
these conditions, the noise reduction is becoming meaningless as
it deforms the analyzed optical field. This process of noise
reduction has been experimentally demonstrated in
\cite{PerinaJr2017b} for a weak twin beam whose photocount
distribution was monitored by an intensified CCD (iCCD) camera.

In this article we report on the development and experimental test of a reconstruction
method for twin beams that takes advantage of the above discussed
noise reduction. We show that the resulting photon-number
distribution reached by the developed method is considerably less
noisy compared to that obtained by the usual approach. For both
reconstructions, we apply the method of maximum likelihood that
has been considered as a workhorse for reconstructions of various
types of quantum states. The idea of the discussed method is presented in Sec.~II.
Quantifiers monitoring the level of noise reduction due to spatial filtering are discussed
in Sec.~III. Sec.~IV is devoted to practical implementation of the method. Conclusions are drawn
in Sec.~V.

\section{Reconstruction of twin-beam photon-number distributions incorporating
spatial noise reduction}

We present the developed method by its comparison with the usual
reconstruction method. In the usual approach, a measured joint
signal-idler photocount histogram $ f(c_{\rm s},c_{\rm i}) $ that
gives the probability of detecting together $ c_{\rm s} $ signal
and $ c_{\rm i} $ idler photocounts from one realization of a twin
beam is directly used in the reconstruction formula. In the
maximum-likelihood approach, the joint signal-idler photon-number
distribution $ p^{\rm std}(n_{\rm s},n_{\rm i}) $ of the
reconstructed twin beam is reached as a steady state of the
following iteration procedure ($ l=0,1,\ldots $)
\cite{Dempster1977,Rehacek2003,PerinaJr2012,Harder2014a}:
\begin{eqnarray} 
 p^{(l+1)}(n_{\rm s},n_{\rm i}) &=& p^{(l)}(n_{\rm s},n_{\rm i})
  \nonumber \\
 & & \hspace{-25mm} \mbox{} \times
  \sum_{c_{\rm s},c_{\rm i}} \frac{ f(c_{\rm s},c_{\rm i})
  T_{\rm s}(c_{\rm s},n_{\rm s}) T_{\rm i}(c_{\rm i},n_{\rm i}) }{
  \sum_{n'_s,n'_i} T_{\rm s}(c_{\rm s},n'_{\rm s})
   T_{\rm i}(c_{\rm i},n'_{\rm i})
   p^{(l)}(n'_{\rm s},n'_{\rm i}) } .
\label{1}
\end{eqnarray}
In Eq.~(\ref{1}), the positive-operator-valued measures (POVMs) $
T_a(c_a,n_a) $ give the probabilities for detecting $ c_a $
photocounts out of $ n_a $ photons in beam $ a $, $ a={\rm s,i} $.
As such they characterize a linear relation between the
(reconstructed) joint signal-idler photon-number distribution $ p
$ and the corresponding experimental photocount histogram $ f $.
This relation is 'inverted' with the help of the formula written
in Eq.~(\ref{1}). The form of POVMs depends on the properties of a
detector. For the used iCCD camera, detection efficiency, mean
dark count number per pixel and number of pixels inside an active
detection area are the parameters entering the formulas for POVMs
$ T_a(c_a,n_a) $ occurring in Eq.~(\ref{1}) [see below in
Eq.~(\ref{9})]. We note that the iteration procedure corrects for
the detrimental effects in detection described explicitly in the
POVMs including the detector noise.

\begin{figure}         
 \centerline{\resizebox{0.8\hsize}{!}{\includegraphics{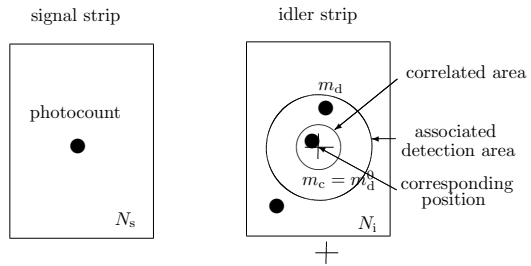}}}
 \vspace{2mm}
 \caption{Typical detection configuration in the signal and idler detection strips on a photocathode of the iCCD camera showing
 one photocount in the signal strip and 3 photocounts in the idler strip. The photocount in the middle of the idler strip
 is positioned within the correlated area (that covers $ m_{\rm c} $ pixels) drawn around the point corresponding
 to the detection position of the photocount in the signal strip and so both photocounts form a photocount pair. For given
 number $ m_{\rm d} $ of pixels in the detection area, the photocount in the upper part of the idler strip lies inside
 the detection area and so it is considered as an unpaired photocount in the reconstruction procedure. Contrary to this,
 the photocount in the lower part of the idler strip lies outside the detection area and so it is identified as a noise unpaired photocount
 and as such it is excluded from the consideration in the reconstruction procedure. The extension of the detection area characterized by
 $ m_{\rm d} $ pixels gradually changes and optimal performance of the reconstruction procedure is expected for
 $ m_{\rm d}^0 \approx m_{\rm c} $. The signal (idler) strip is composed of $ N_{\rm s} $ ($ N_{\rm i} $) pixels.}
\label{fig1}
\end{figure}

Contrary to this, the suggested method exploits filtering based on
spatial correlations \cite{PerinaJr2017b} to arrive at a histogram
$ f_{\rm p}(c_{\rm p};m_{\rm d}) $ of paired signal and idler
photocounts and a joint histogram $ f_{\rm si}^{\rm unp}(c_{\rm
s},c_{\rm i};m_{\rm d}) $ of unpaired photocounts that occur in
the neighborhood (inside the detection area) of the identified photocount
pairs. Details of identification of individual photocounts and different types of photocount
configurations in both signal and idler detection strips
are found in the caption to Fig.~\ref{fig1}.
Both histograms $ f_{\rm p}(c_{\rm p};m_{\rm d}) $ and $ f_{\rm si}^{\rm unp}(c_{\rm
s},c_{\rm i};m_{\rm d}) $ depend on the extension of the detection
area that, in case of CCD detection elements, is conveniently parameterized
by the number $ m_{\rm d} $ of pixels inside this area.
Reconstruction of both fields described by the histograms $ f_{\rm
p}(c_{\rm p};m_{\rm d}) $ and $ f_{\rm si}^{\rm unp}(c_{\rm
s},c_{\rm i};m_{\rm d}) $ is based upon the iteration procedure of
Eq.~(\ref{1}) (and its one-dimensional variant) in which the
noiseless POVMs $ T_a^{\rm nl}(c_a,n_a) $ are used [see Eq.~(\ref{9}) below, $ D = 0 $]. We note that
the noiseless POVMs are applied as the noise is assumed to be
(partially) removed by filtering via spatial correlations.
The usual detection efficiencies $ \eta_{\rm s} $ and $ \eta_{\rm
i} $ appropriate for the signal and idler beam (detection strip), respectively, are
used in reconstructing the joint signal-idler photon-number
distribution $ p_{\rm si}^{\rm unp}(n_{\rm s},n_{\rm i};m_{\rm
d}) $ from the histogram $ f_{\rm si}^{\rm unp}(c_{\rm s},c_{\rm
i};m_{\rm d}) $ of unpaired photocounts. Contrary to this, an effective detection
efficiency $ \eta_{\rm p}(m_{\rm d}) $ has to be applied when
reconstructing the distribution $ p_{\rm p} $ of photon pairs from
the histogram $ f_{\rm p}(c_{\rm p};m_{\rm d}) $. Provided that
the positions of signal photocounts in the signal detections strip are used to define the
overall detection area in the idler detection strip that is taken into account for
the spatial filtering, the effective detection efficiency $
\eta_{\rm i,p}^{\rm eff} $ is given as (for more details, see \cite{PerinaJr2017b})
\begin{equation} 
 \eta_{\rm i,p}^{\rm eff}(m_{\rm d}) = \frac{\langle c_{\rm i} \rangle^{\rm
 red}(m_{\rm d}) }{ \langle c_{\rm i} \rangle } \eta_{\rm s},
\label{2}
\end{equation}
where $ \langle c_{\rm i} \rangle^{\rm red}(m_{\rm d}) $ [$
\langle c_{\rm i} \rangle $] gives the mean number of idler
photocounts considered with [without] spatial filtering. For the
detection area with $ m_{\rm d} $ pixels, an overall joint
signal-idler photon-number distribution $ p(n_{\rm
s},n_{\rm i};m_{\rm d}) $ is obtained by the following
convolution:
\begin{eqnarray}  
 p(n_{\rm s},n_{\rm i};m_{\rm d}) &=& \sum_{n_{\rm p}=0}^{
  \min (n_{\rm s},n_{\rm i}) } p_{\rm p}(n_{\rm p};m_{\rm d}) \nonumber \\
 & & \mbox{} \hspace{-10mm} \times p_{\rm si}^{\rm unp}(n_{\rm s}-n_{\rm p},
 n_{\rm i}-n_{\rm p};m_{\rm d})
\label{3}
\end{eqnarray}
and the distribution $ p_{\rm p} $ is derived from the histogram
$ f_{\rm p} $. Proper choice of the number $ m_{\rm d}^0 $ of pixels in the
detection area then identifies the joint photo-number distribution
$ p(n_{\rm s},n_{\rm i};m_{\rm d}^0) $ of the
reconstructed twin beam.

\section{Quantifiers useful for monitoring the level of noise reduction}

To monitor the performance of the noise reduction in the suggested reconstruction method,
we need to quantify (quantum) correlations between the signal and
idler fields both for the experimental photocounts and
reconstructed photon numbers. Here, we show that the
non-classicality depths $ \tau $ \cite{Lee1991} determined for the
non-classicality identifiers $ E_k $, $ k=2,3,4 $, defined in
terms of intensity moments $ \langle W^l \rangle $ as
\cite{PerinaJr2017a}
\begin{eqnarray}   
 E_2 &=& \langle W_{\rm s}^2\rangle + \langle W_{\rm i}^2 \rangle - 2 \langle W_{\rm s}
  W_{\rm i}\rangle,  \label{4} \\
 E_3 &=& \langle W_{\rm s}^3\rangle + \langle W_{\rm i}^3 \rangle - \langle W_{\rm s}^2
  W_{\rm i}\rangle - \langle W_{\rm s} W_{\rm i}^2\rangle, \label{5} \\
 E_4 &=& \langle W_{\rm s}^4\rangle + \langle W_{\rm i}^4 \rangle - 2\langle W_{\rm s}^2
  W_{\rm i}^2\rangle  \label{6}
\end{eqnarray}
allow for relatively precise determination of the number $ m_{\rm
d}^0 $ of detection pixels in the detection area that leads to the best result
of the reconstruction procedure. We note that an $ l $-th intensity
moment $ \langle W^l \rangle $ is related to the moments $ \langle
c^k \rangle $ of photocounts by the formula $ \langle W^l \rangle
= \sum_{k=1}^l S_{lk}^{-1} \langle c^k \rangle $ that uses the
Stirling numbers $ S_{lk} $ of the second kind. We remind that a
non-classicality depth $ \tau $ is given by the number of thermal
photons needed to conceal nonclassical properties of an optical
field visible in a given non-classicality identifier
\cite{PerinaJr2017a}. For comparison, we also determine the
traditional covariance $ C_{\Delta c} $ of fluctuations of the
numbers $ c_{\rm s} $ and $ c_{\rm i} $ of the signal and idler
photocounts and sub-shot-noise parameter $ R_c $,
\begin{eqnarray}   
 C_{\Delta c} &=& \frac{ \langle \Delta c_{\rm s} \Delta c_{\rm i} \rangle}{
  \sqrt{\langle (\Delta c_{\rm s})^2 \rangle \langle (\Delta c_{\rm i})^2 \rangle} },
\label{7} \\
 R_c &=& \frac{ \langle [\Delta(c_{\rm s}-c_{\rm i})]^2\rangle}{ \langle c_{\rm s}
  \rangle + \langle c_{\rm i} \rangle } .
\label{8}
\end{eqnarray}

\section{Experimental implementation of the developed reconstruction method}

\begin{figure}         
 \centerline{\resizebox{0.8\hsize}{!}{\includegraphics{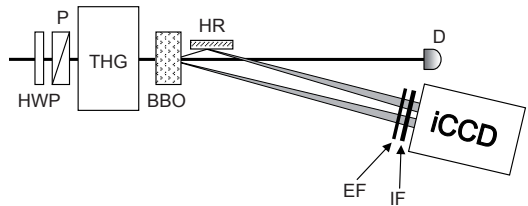}}}
 \vspace{2mm}
 \caption{Scheme of the experiment. The third harmonics (THG, 280 nm) of a Ti:sapphire laser beam pumps a BaB$ {}_2 $O$ {}_4 $
  (BBO) nonlinear crystal. Nearly degenerate signal and idler (steered by high-reflectivity mirror HR)
  beams are filtered by a 14-nm-wide bandpass frequency filter IF and then detected
  in two detection strips on a photocathode of iCCD camera. Long-pass (above 490 nm) filter EF diminishes the noise.
  Intensity of the pump beam monitored by detector D is actively stabilized (rms below 0.3\%) using
  motorized half-wave plate HWP followed by polarizer P. According
  to the calibration procedure \cite{PerinaJr2012a} applied to the experimental data,
  the signal (idler) detection strip was composed of $ N_{\rm s} = N_{\rm i} = 6500 $ pixels,
  exhibited detection efficiency $ \eta_{\rm s} = 0.228 \pm 0.005 $
  ($ \eta_{\rm i} = 0.223 \pm 0.005 $) and suffered by $ D_{\rm s} = 0.2/N_{\rm s} $
  ($ D_{\rm i} = 0.2/N_{\rm i} $) mean dark count number per pixel.}
\label{fig2}
\end{figure}

To compare the performance of the suggested reconstruction method with
the standard one, we measured a joint signal-idler photocount
histogram $ f $ of a twin beam centered at the wavelength 560~nm and
originating in a non-collinear type-I interaction in a 5-mm long
BaB$ {}_2 $O$ {}_4 $ crystal pumped by the third harmonics of a
femtosecond cavity dumped Ti:sapphire laser (pulse duration
150~fs, central wavelength 840~nm) \cite{Hamar2010}. The signal
and idler photocounts were captured in different detection strips
on a photocathode of an iCCD camera Andor DH334-18U-63 (for the
geometry of experimental setup, see Fig.~\ref{fig2}); $ 1.2 \times
10^6 $ experimental repetitions were performed.
Provided that a detection strip is composed of $ N $ pixels, has
detection efficiency $ \eta $ and its mean dark count number per pixel is
$ D $, its POVM $ T $ needed in Eq.~(\ref{1}) for the
reconstruction procedure is written as \cite{PerinaJr2012}:
\begin{eqnarray}     
  T(c,n) &=& \left(\begin{array}{c} N \\ c \end{array}\right) (1-D)^{N}
   (1-\eta)^{n} (-1)^{c} \nonumber \\
  & &  \mbox{} \hspace{-15mm} \times  \sum_{l=0}^{c} \left(\begin{array}{c} c \\ l \end{array}\right)
    \frac{(-1)^l}{(1-D)^l}  \left( 1 + \frac{l}{N} \frac{\eta}{1-\eta}
   \right)^{n}.
\label{9}
\end{eqnarray}

\begin{figure}  
 \centerline{(a)\hspace{3mm}\resizebox{0.87\hsize}{!}{\includegraphics{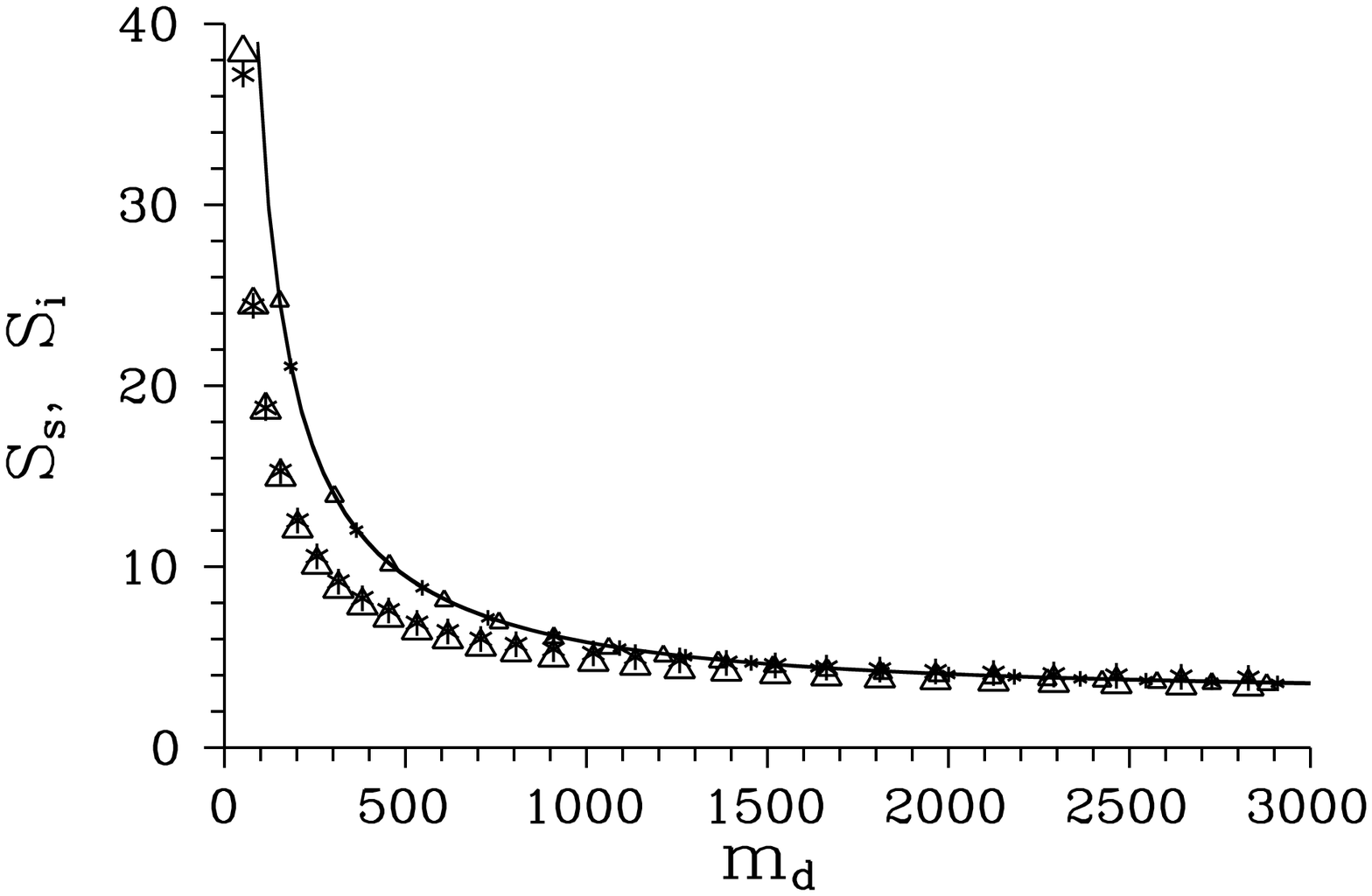}}}
  \vspace{2mm}
 \centerline{(b)\hspace{3mm}\resizebox{0.87\hsize}{!}{\includegraphics{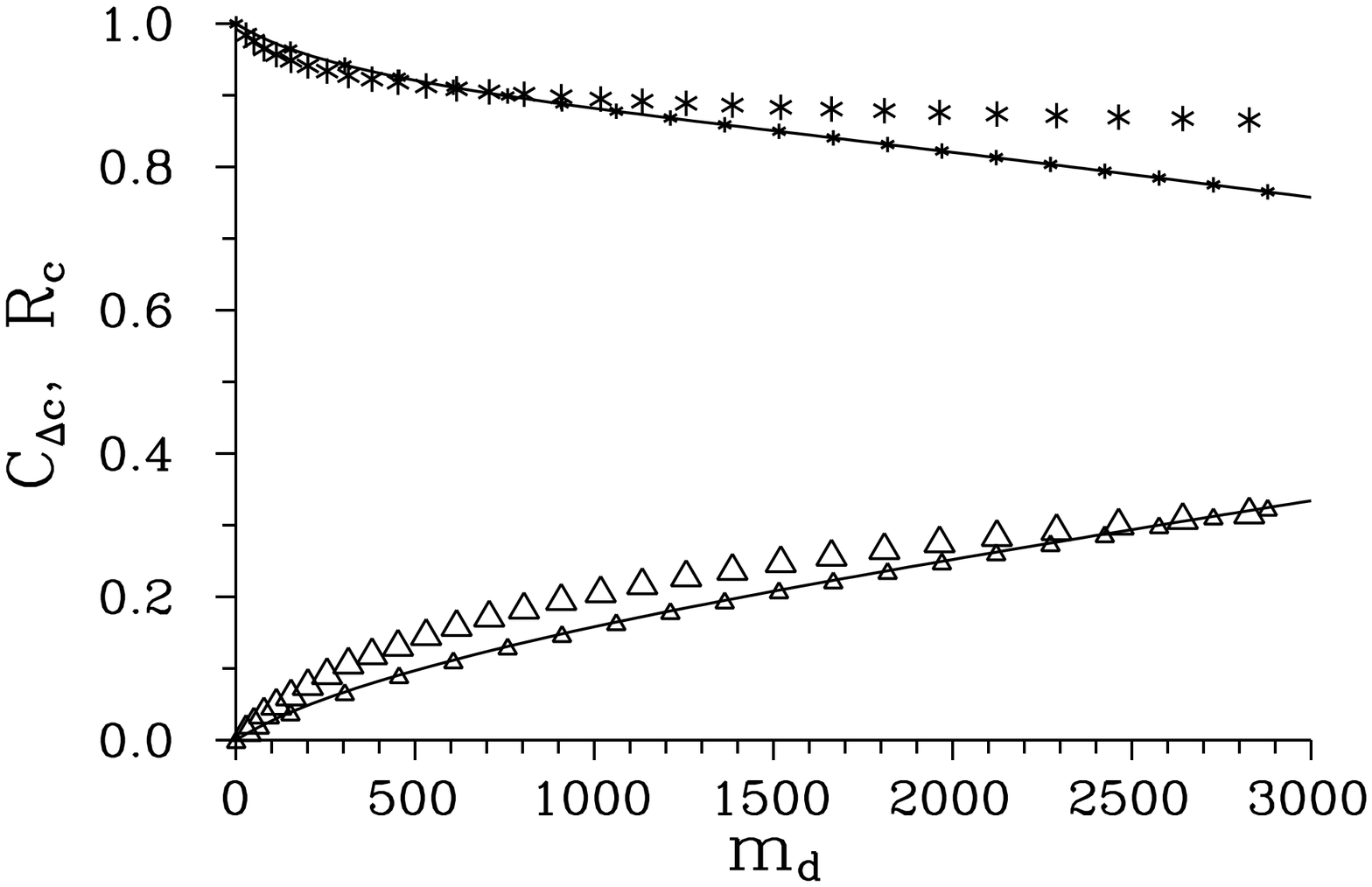}}}
  \vspace{2mm}
 \centerline{(c)\hspace{3mm}\resizebox{0.87\hsize}{!}{\includegraphics{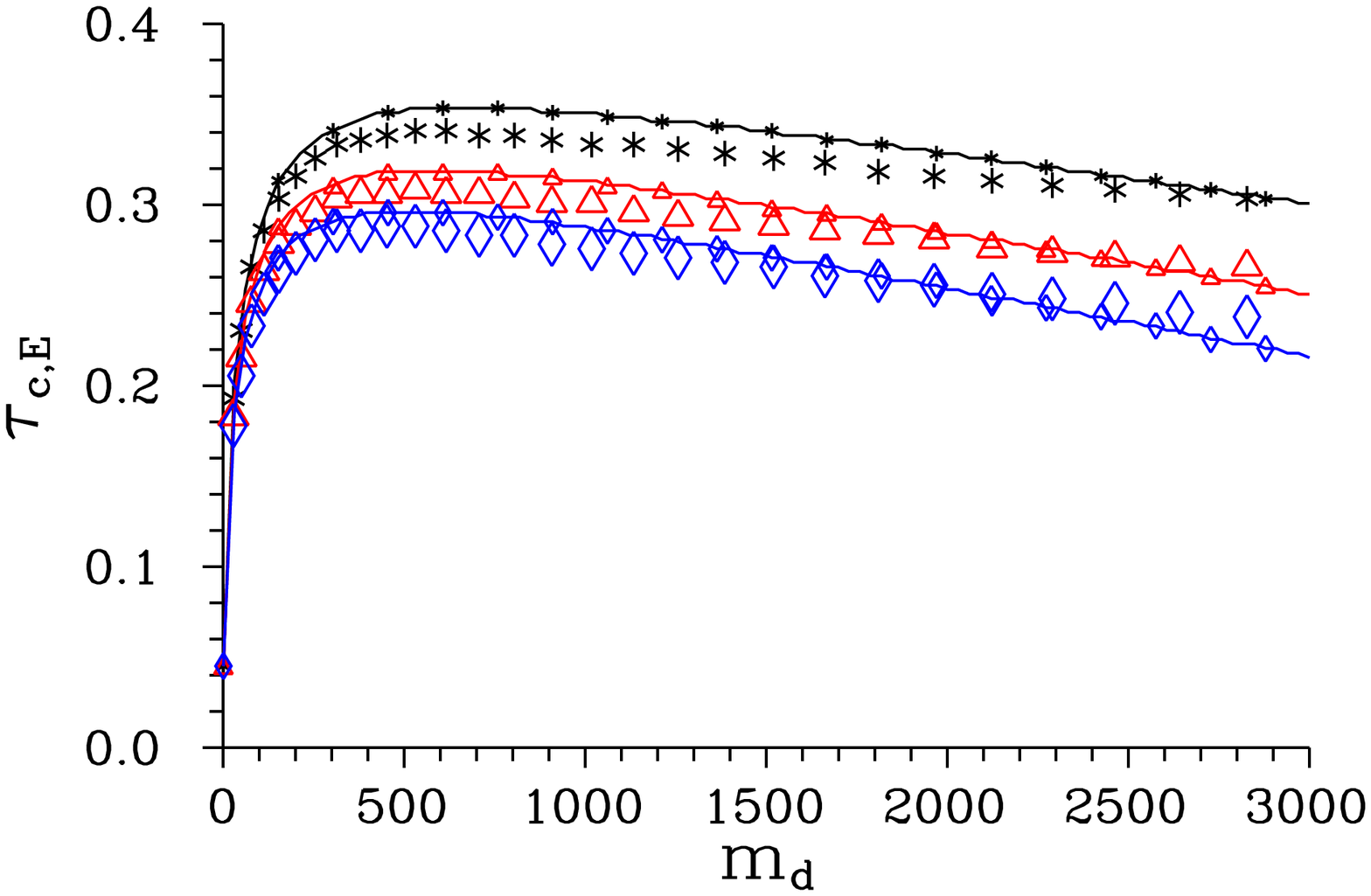}}}
  \vspace{2mm}
 \caption{(a) Signal-to-noise ratios $ S_{a} \equiv \langle c_p\rangle / \langle c_a\rangle $ for
  $ a={\rm s} $ ($ \ast $) and $ a={\rm i} $ ($ \triangle $), (b)
  covariance $ C_{\Delta c} $ ($ \ast $) and sub-shot-noise parameter $ R_c
  $ ($ \triangle $) and (c) non-classicality depths $ \tau_{c,E_k}
  $ for $ k=2 $ ($ \ast $), 3 (red $ \triangle $) and 4 (blue $ \diamond $) determined from
  the photocount measurement as they depend on the number $ m_{\rm d} $ of pixels in the
  detection area. Experimental values are plotted as isolated
  symbols, the corresponding solid curves were drawn following the standard model
  of Ref.~\cite{PerinaJr2017b}. Relative experimental errors for
  $ S_{\rm s} $, $ S_{\rm i} $, $ C_{\Delta c} $, $ R_c $, $
  \tau_{c,E_2} $, $ \tau_{c,E_3} $, and $ \tau_{c,E_4} $ are
  lower than in turn 1\%, 1\%, 3\%, 3\%, 2\%, 3\% and 4\%.}
 \label{fig3}
\end{figure}

In the suggested method, reduction of the noise in the
experimental data is achieved by spatial filtering of the photocounts whose strength
gradually increases with the decreasing number $ m_{\rm d} $ of
pixels in the considered detection areas drawn around the identified photocount pairs.
This leads both to the
decrease of the number $ \langle c_p\rangle $ of identified
photocount pairs as well as the decrease of the numbers $ \langle
c_{\rm s}\rangle $ and $ \langle c_{\rm i}\rangle $ of the signal
and idler photocounts found inside these detection areas. The essence
of the suggested method is based on the fact that the numbers $
\langle c_{\rm s}\rangle $ and $ \langle c_{\rm i}\rangle $
decrease relatively faster than the number $ \langle c_p\rangle $
with decreasing number $ m_{\rm d} $ of detection pixels. So, the signal-to-noise ratios $
S_{a} \equiv \langle c_p\rangle / \langle c_a\rangle $, $ a={\rm
s,i} $, increase with decreasing $ m_{\rm d} $. For the analyzed
experimental data and following the curves in Fig.~\ref{fig3}(a),
we have $ S_a \approx 10 $, $ a={\rm s,i} $, for $ m_{\rm d}^0 =
290 $ compared to $ S_a \approx 4 $ determined without spatial
filtering. We note that, for our experimental data, the detection
area with $ m_{\rm d}^0 = 290 $ pixels just covers the correlated
area whose profile can also be deduced from the obtained data (for
details, see \cite{PerinaJr2017b}). Thus, the signal-to-noise
ratios are improved more than two times by the filtering.

Spatial filtering improves correlations between the signal and
idler photocount numbers. In case of the covariance $ C_{\Delta c}
$ of the fluctuations of the signal and idler photocount numbers
and the corresponding sub-shot-noise parameter $ R_c $ defined in
Eqs.~(\ref{7}) and (\ref{8}), respectively, these quantities tend
to reach their optimal values for the negligibly small detection
area ($ C_{\Delta c} \rightarrow 1 $, $ R_c \rightarrow 0 $ for $
m_{\rm d} \rightarrow 0 $), as documented in Fig.~\ref{fig3}(b).
In contrast to this and according to the curves of
Fig.~\ref{fig3}(c), the non-classicality depths $ \tau_{c,E_k} $
belonging to the non-classicality identifiers $ E_k $, $ k=2,3,4
$, from Eqs.~(\ref{4}---\ref{6}) reach their maximal values in the
range of $ m_{\rm d} $ where the detection area approximately
coincides with the correlated area. The values of $ \tau_{c,E_k} $
rapidly drop down to zero when the detection area becomes smaller
than the correlated area. For this reason, the non-classicality
depths $ \tau_{c,E_k} $ are qualitatively better for monitoring
the process of spatial filtering. Also, the non-classicality
identifiers $ E_2 $, $ E_3 $ and $ E_4 $ that are in turn based on
the second-, third- and fourth-order intensity moments behave
similarly in the process of spatial filtering. This indicates that
the spatial filtering systematically modifies correlations of
different orders.

\begin{figure}  
 \centerline{\resizebox{0.8\hsize}{!}{\includegraphics{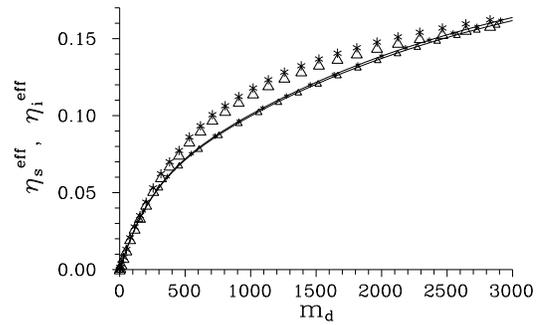}}}
  \vspace{2mm}
 \caption{Effective efficiencies $ \eta_{\rm s,p}^{\rm eff} $ ($ \ast
  $) and $ \eta_{\rm i,p}^{\rm eff} $ ($ \triangle $) defined in
  Eq.~(\ref{2}) as they depend on the number $ m_{\rm d} $ of pixels in the
  detection area. Experimental values are plotted as isolated
  symbols, the corresponding solid curves were drawn following the standard model
  of Ref.~\cite{PerinaJr2017b}. Relative experimental errors are
  better than 2\%.}
 \label{fig4}
\end{figure}

Now let us have a look how these modifications in the experimental
photocount histograms affect the behavior of correlations in the
reconstructed photon-number distributions. The determination of
the effective detection efficiency $ \eta_{\rm i,p}^{\rm eff} $
(or $ \eta_{\rm s,p}^{\rm eff} $) that is used for reconstructing
the paired part of the photocount field represents the most important
(though technical) step in the whole reconstruction. These effective efficiencies naturally
decrease with the decreasing number $ m_{\rm d} $ of detection pixels and they attain values
around $ \eta_{\rm s} \eta_{\rm i} $ when the detection area is
close to the correlated area (see Fig.~\ref{fig4}). These values
are then used to arrive at the proper photon-number distribution
of the analyzed twin beam.

\begin{figure}  
 \centerline{(a)\hspace{3mm}\resizebox{0.8\hsize}{!}{\includegraphics{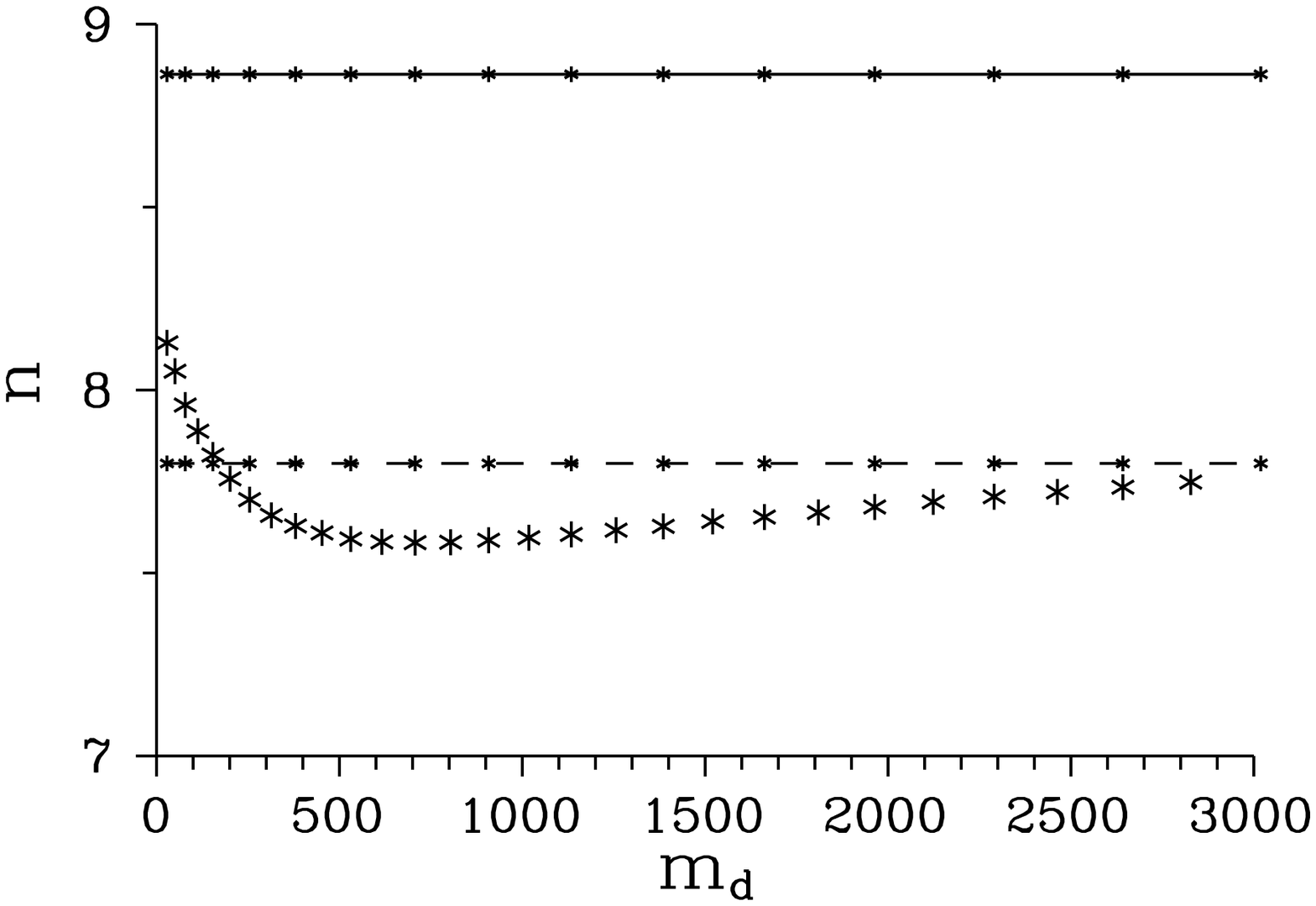}}}
  \vspace{2mm}
 \centerline{(b)\hspace{3mm}\resizebox{0.87\hsize}{!}{\includegraphics{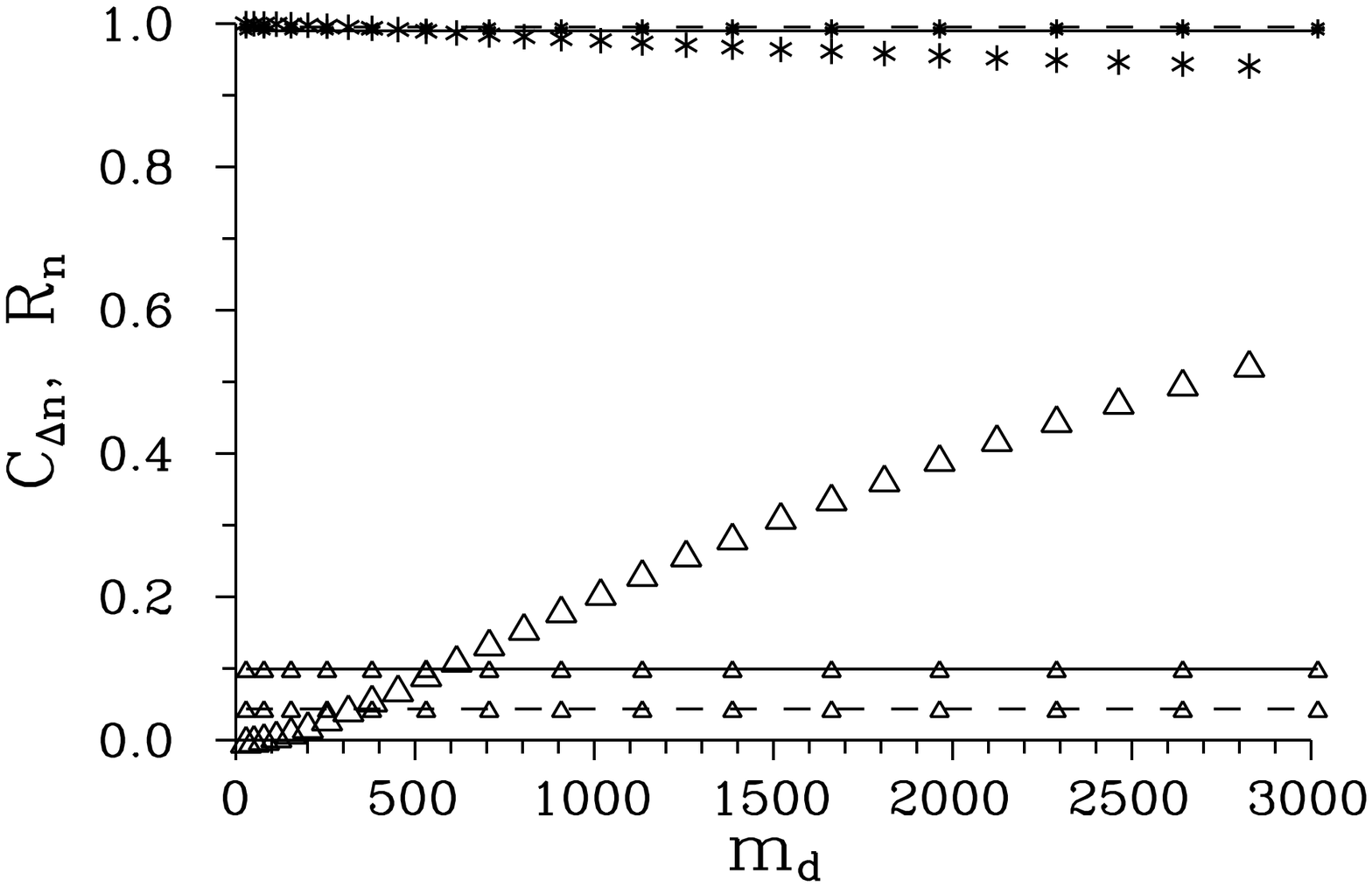}}}
  \vspace{2mm}
 \centerline{(c)\hspace{3mm}\resizebox{0.87\hsize}{!}{\includegraphics{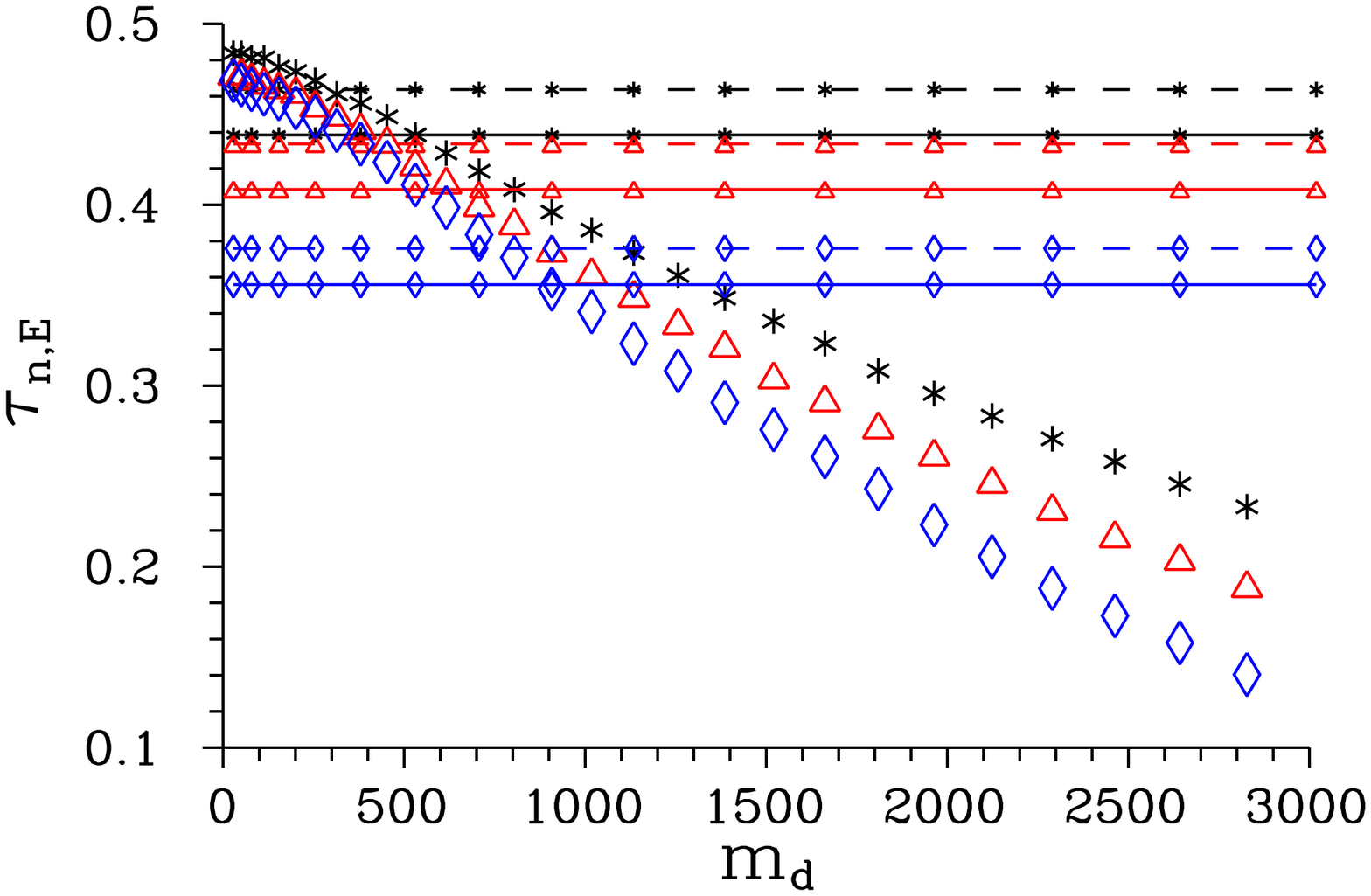}}}
  \vspace{2mm}
 \caption{(a) Mean photon number $ \langle n\rangle \equiv (\langle n_{\rm s}\rangle +
  \langle n_{\rm i}\rangle )/2 $ ($ \ast $), (b) covariance $ C_{\Delta n} $ ($ \ast $) and
  sub-shot-noise parameter $ R_n $ ($ \triangle $) and (c) non-classicality depths $ \tau_{n,E_k}
  $ for k=2 ($ \ast $), 3 (red $ \triangle $) and 4 (blue $ \diamond $) determined
  for the reconstructed twin beam are plotted as isolated symbols for different numbers $ m_{\rm d} $ of pixels in the
  detection area. Horizontal lines give the values reached by the
  standard reconstruction method with POVMs including the intrinsic detector
  mean dark count numbers ($ D_a N_a = 0.2 $, $ a={\rm s,i} $; solid
  lines) and suitably increased mean dark count numbers ($ D_a N_a = 0.45 $, $ a={\rm s,i} $;
  dashed lines). Relative experimental errors for
  $ \langle n\rangle $, $ C_{\Delta n} $, $ R_n $, $
  \tau_{n,E_2} $, $ \tau_{n,E_3} $, and $ \tau_{n,E_4} $ are
  lower than in turn 1\%, 3\%, 3\%, 2\%, 3\% and 4\%.}
 \label{fig5}
\end{figure}

Properties of a joint signal-idler photon-number distribution $ p
$ reached by the reconstruction method depend on the strength of
the spatial filtering, as documented in Fig.~\ref{fig5} where the
most important quantities of the reconstructed twin beam are
plotted as functions of the number $ m_{\rm d} $ of pixels in the
detection areas.
According to the curves of Fig.~\ref{fig5}(a), the mean
photon-number $ \langle n \rangle $ [$ \langle n\rangle = (\langle
n_{\rm s}\rangle + \langle n_{\rm i}\rangle )/2 $] of the
reconstructed twin beam slightly decreases with the increasing
spatial filtering. However, it starts to increase when the
detection area is comparable to the correlated area. The initial
decrease of the mean photon-number $ \langle n \rangle $ with
decreasing number $ m_{\rm d} $ observed for greater numbers $
m_{\rm d} $ has two reasons. First, the unwanted noise present in
the experimental data is gradually removed. Second, the number $
\langle c_{\rm p}\rangle (m_{\rm d}) $ of identified photocount
pairs is greater than the actual one due to the existence of
accidental photocount pairs (for details, see
\cite{PerinaJr2017b}) on the one hand, on the other hand the
corresponding effective detection efficiency $ \eta_{\rm s,p}^{\rm
eff} $ that is constructed to compensate for the effect of
accidental pairing is slightly overestimated. In our experiment,
both contributions are comparably strong. Whereas the first effect
just demonstrates the essence of the spatial noise reduction, the
second effect distorts the experimental data and as such it is
unwanted. However, this effect is minimal (and ideally disappears)
when the detection area just covers the correlated area. In this
case, the number of accidental photocount pairs is already very
low and, according to the theory of absolute detector calibration
\cite{Klyshko1980}, the definition (\ref{2}) of the effective
detection efficiency $ \eta_{\rm i,p}^{\rm eff} $ gives us $
\eta_{\rm i,p}^{\rm eff} \approx \eta_{\rm s}^2 $. The appropriate
number $ m_{\rm d}^0 $ of pixels in the detection area is ideally
revealed by determining the correlated area (or even more
precisely by determining the profile of the correlated area
\cite{PerinaJr2017b}).

The reconstructed fields are naturally endowed with better covariances
$ C_{\Delta n} $ of the fluctuations of photon numbers compared to the original
covariances $ C_{\Delta c} $ of the fluctuations of photocount numbers, as evidenced
by comparing the graphs in Figs.~\ref{fig3}(b) and \ref{fig5}(b). On the other hand, direct
comparison of sub-shot-noise parameters $ R_n $ and $ R_c $ as well as non-classicality depths
$ \tau_{n,E_k} $ and $ \tau_{c,E_k} $ for $ k=2,3,4 $ belonging both to photocount and photon-number
fields is not possible as the reconstructed fields are roughly four times more intense.
On the other hand, mutual comparison of the curves for non-classicality depths
$ \tau_{c,E_k} $ and $ \tau_{n,E_k} $ drawn in Figs.~\ref{fig3}(c) and \ref{fig5}(c)
reveals qualitatively different behavior of these depths in the area of $ m_{\rm d} $ where the
detection area is comparable or smaller than the correlated area. This difference
comes from the fact that the reconstruction method treats differently photocount pairs and individual
noisy (unpaired) photocounts. Roughly speaking the photocount pairs are amplified stronger (as $ 1/ \eta_{\rm s,p}^{\rm eff} $)
than the individual noisy photocounts (as $ 1/ \eta_{\rm s} $) in the reconstruction method.
We note that, in our experiment, we have $ \eta_{\rm s,p}^{\rm eff}(m_{\rm d}^0 =290) \approx
0.04 $ compared to $ \eta_{\rm s} \approx 0.2 $.

\begin{figure}  
 \centerline{(a)\hspace{3mm}\resizebox{0.9\hsize}{!}{\includegraphics{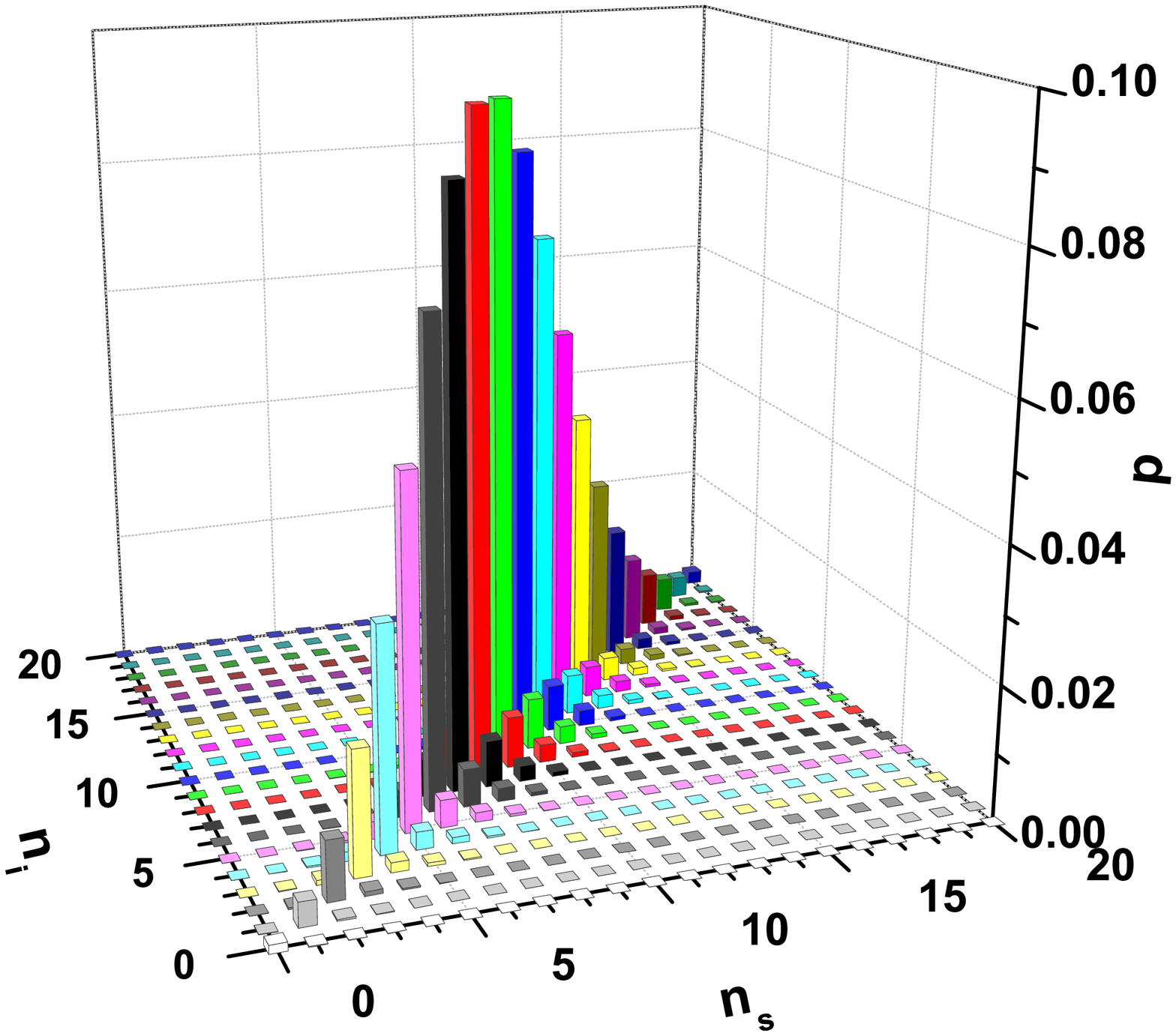}}}
  \vspace{2mm}
 \centerline{(b)\hspace{3mm}\resizebox{0.9\hsize}{!}{\includegraphics{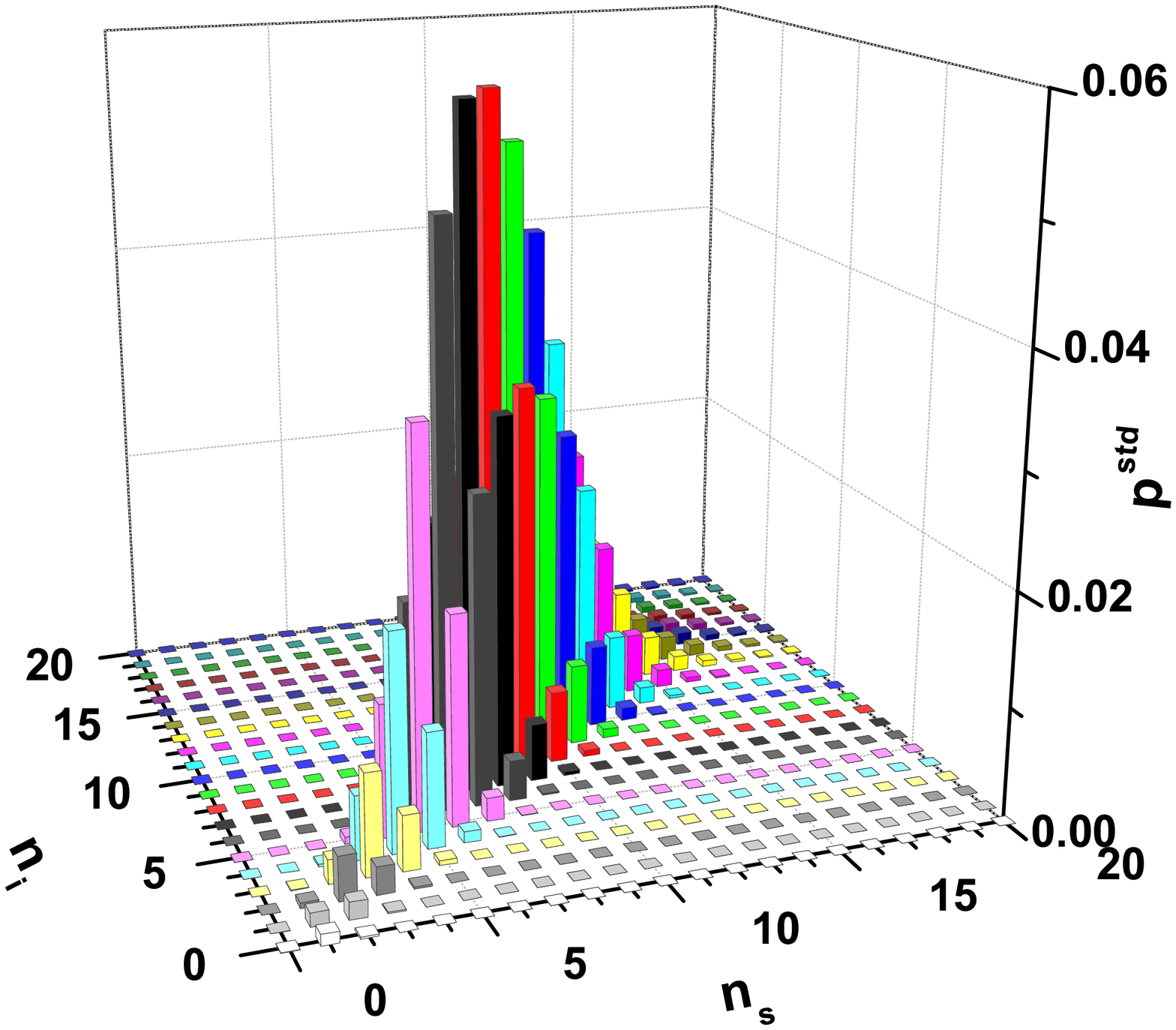}}}
  \vspace{2mm}
 \caption{Joint signal-idler photon-number distribution $ p $ of the reconstructed twin beam reached (a)
  with the help of filtering by spatial correlations and (b) by the standard reconstruction procedure.}
 \label{fig6}
\end{figure}

The analysis of the experimental data revealed the correlated area with 290 pixels
(for details, see \cite{PerinaJr2017b}). The reconstructed joint signal-idler photon-number distribution $ p $ obtained
for $ m_{\rm d}^0 =290 $ pixels in the detection area, that we consider as the best
from the point of view of the developed method, is shown in Fig.~\ref{fig6}(a).
For comparison, we draw in Fig.~\ref{fig6}(b) a photon-number distribution
$ p^{\rm std} $ reached by the standard reconstruction method that includes
the intrinsic detector mean dark count numbers $ D_{\rm s} $ and $ D_{\rm i} $ as specified in the caption to
Fig.~\ref{fig2}.
Values of the parameters $ \langle n \rangle $, $ C_{\Delta n} $, $ R_n $ and $ \tau_{n,E_k} $,
$ k=2,3,4 $, of this standardly reconstructed twin beam, that are plotted in Figs.~\ref{fig5}(a)---(c) by solid horizontal lines,
show that the mean photon-number $ \langle n \rangle $ is about 15\% larger and the noise in this twin beam
is also larger compared to the twin beam revealed by the developed reconstruction method. The greater
mean photon-number $ \langle n \rangle $ indicates that considerable amount of the optical noise
is present in the original (i.e. not spatially filtered) experimental photocount histogram. This amount of optical noise can phenomenologically
be removed in the standard reconstruction procedure by considering greater (effective) values of the
mean dark count numbers $ D_{\rm s} $ and $ D_{\rm i} $. To demonstrate the performance of this approach, we have plotted
by dashed horizontal lines in the graphs of Figs.~\ref{fig5}(a)---(c) the values of parameters appropriate for
$ D_{\rm s}N_{\rm s} = D_{\rm i}N_{\rm i} = 0.45 $. The chosen amount of the optical noise approximately leads to the correct
mean photon-number $ \langle n \rangle $, but the values of non-classicality depths $ \tau_{n,E_k} $, $ k=2,3,4 $, remain
worse in comparison with those characterizing the reconstructed twin beam in Fig.~\ref{fig6}(a). Detailed analysis of the curves plotted
in Fig.~\ref{fig5}(c) reveals that the greater the power of intensity moments involved in the determination of the non-classicality
depth $ \tau_{n,E_k} $, the worse the noise reduction by the standard reconstruction procedure compared to the developed method.
These results clearly show that the suggested reconstruction method exploiting filtering by spatial correlations is superior above
the standard one.

\section{Conclusions}

We have suggested and elaborated a method for reconstructing joint photon-number distributions of
twin beams from the measured photocount histograms that uses spatial filtering of the experimental photocounts to reduce
the experimental noise. The applied filtering exploits spatial correlations of photons in a twin beam.
In the experimental implementation, we demonstrated superior performance of the developed reconstruction
method above the standard one. Though the processing of experimental data in the developed method is considerably more involved
in comparison with the standard approach, the developed method brings considerable advantages and we suggest its application
wherever the spatially resolved photocount histograms of twin beams are at disposal.

\section{Acknowledgments} The authors thank M. Hamar for his
help with the experiment. O.H. and V.M. were supported by the GA
\v{C}R project 18-08874S. J.P. was supported by the GA
\v{C}R project 18-22102S.


%

\end{document}